\documentclass[pre,twocolumn,amsmath,amssymb,floatfix]{revtex4}

\usepackage{graphicx}
\usepackage{dcolumn} 
\usepackage{bm}      
\usepackage{color}

\begin{document}

\title{Thermalization and free decay in Surface Quasi-Geostrophic flows}
\author{Tomas Teitelbaum$^{1}$ and Pablo D.~Mininni$^{1,2}$}
\affiliation{$^1$ Departamento de F\'\i sica, Facultad de Ciencias 
Exactas y Naturales, Universidad de Buenos Aires and IFIBA, CONICET, 
Ciudad Universitaria, 1428 Buenos Aires, Argentina. \\
             $^2$ NCAR, P.O. Box 3000, Boulder, Colorado 80307-3000, U.S.A.}
\date{\today}

\begin{abstract}
We derive statistical equilibrium solutions of the truncated inviscid surface 
quasi-geostrophic (SQG) equations, and verify the validity of these solutions 
at late times in numerical simulations. The results indicate the 
pseudo-enstrophy thermalizes while the pseudo-energy can condense at the 
gravest modes, in agreement with previous indications of a direct cascade 
of pseudo-enstrophy and an inverse cascade of pseudo-energy in 
forced-dissipative SQG systems. At early times, the truncated inviscid 
SQG simulations show a behavior reminiscent of forced-dissipative SQG 
turbulence, and we identify spectral scaling laws for the pseudo-energy and 
pseudo-enstrophy spectra. More importantly, a comparison between 
viscous and inviscid simulations allows us to identify free-decay laws 
for the pseudo-enstrophy in SQG turbulence at very large Reynolds number.
\end{abstract}
\maketitle

\section{\label{sec:Intro}Introduction}

Classical Gibbs ensemble methods have been extensively applied to Galerkin 
representation of turbulent systems, and many examples can be found in the 
literature. The first studies of the statistical mechanics of discrete 
distributions of vortices in a two-dimensional (2D) flow using a Hamiltonian 
formalism can be found in \cite{Onsager1949}. Later, it was shown 
\cite{Lee1952} that Gibbsian statistical mechanics can be applied to 
Galerkin truncations of the hydrodynamic and magneto hydrodynamic (MHD) 
equations. This allowed many studies of continuous vorticity distribution in 
truncated 2D inviscid flows 
\cite{Kraichnan1967,Fox1973,Kraichnan1975,Kraichnan1980}, where absolute 
equilibrium spectra for the quadratic conserved quantities of the system 
were derived. Following \cite{Kraichnan1967}, in \cite{Kraichnan1973} the 
absolute inviscid equilibrium ensemble for three-dimensional (3D) Euler 
flows was considered. Later, 3D inviscid magneto hydrodynamic MHD 
equilibrium solutions were investigated in \cite{Stribling1990}, 
and recently the approach was extended to Hall-MHD in \cite{Servidio2008}. 
Other problems studied in this framework include geophysical flows 
\cite{Waite2004}, fast rotating flows \cite{Borouiba2008,Mininni2011}, and 
formulations of one- and two-layer quasi-geostrophic models 
\cite{Salmon1976}. In general, classical Gibbs ensemble methods can be 
applied to systems with quadratic conserved quantities and that satisfy the 
Liouville theorem, which implies the incompressibility of the flow in phase 
space (e.g., the space of complex Fourier modes).

In the absolute equilibrium, fields have Gaussian statistics and the quadratic 
conserved quantities of the system have associated temperatures that can 
be positive or negative. In the former case, the quantity is said to 
``thermalize'', and the quantity in the equilibrium is equally distributed 
among all modes in the system. In the latter case the quantity ``condenses,'' 
and is accumulated at the gravest modes. However, forced-dissipative turbulent 
systems are far from equilibrium, as the effect of viscosity prevents 
relaxation towards a true equilibrium state, giving rise to solutions with 
non-zero flux (cascades) and making direct comparison with equilibrium 
solutions inadequate. In spite of this, Gibbs ensemble methods proved 
useful to predict the direction of the cascades in many forced-dissipative 
systems, depending on whether the quantity of interest thermalized or 
condensed in the associated truncated inviscid system (see, e.g., 
\cite{Fox1973,Kraichnan1973,Frisch1975,Waite2004}, although note not 
all inverse cascades in forced-dissipative systems are associated with 
condensates in the statistical equilibrium of the equivalent truncated 
ideal system \cite{Borouiba2008,Mininni2011}).

The recent finding of transients and quasi-stable states in truncated 
inviscid flows, in which non-thermalized modes behave as in the viscous 
case (see, e.g., 
\cite{Cichowlas2005,Krstulovic2008,Frisch2008,Krstulovic2009}), has 
renewed interest in classical Gibbs ensembles, indicating more 
information than just the direction of the cascades can be extracted 
from these systems. In particular, it was found that at early times and 
as the system evolves towards equilibrium, a comparison between the 
inviscid system and a viscous turbulent flow can be achieved by 
considering in the inviscid system the net effect of the modes that 
have already thermalized as an effective viscosity acting on the 
non-thermalized modes \cite{Krstulovic2008,Krstulovic2009}. This 
viscous-like dynamics was reported in many systems, including the 3D 
truncated Euler equation \cite{Cichowlas2005,Krstulovic2008,Krstulovic2009}, 
3D truncated rotating flows \cite{Mininni2011}, 2D truncated MHD 
\cite{Krstulovic2011a}, and turbulent Bose-Einstein condensates using the 
truncated Gross-Pitaevskii equations \cite{Krstulovic2011b}. In this 
latter case, the truncated equations allowed the study of thermalization 
of Bose-Einstein condensates at finite temperature.

Classical Gibbs ensembles and quasi-stable states were also used to study 
the behavior of atmospheric models (see, e.g., \cite{Majda2006}), as the 
need to study atmospheric and oceanic flow dynamics has led to a variety of 
approximate models derived from the 3D Navier-Stokes equation for 
stratified and rotating flows (see for example 
\cite{Charney1948,Charney1971,Pedlosky1987}). In the often used 
geostrophic approximation, the vertical component of the velocity 
field is assumed zero ($u_z=0$), and hydrostatic balance is solved in that 
direction while a linear balance between the Coriolis force and the pressure 
gradient is solved on the horizontal plane. Another family of models is 
given by the so-called quasi-geostrophic (QG) models, which are a first 
order departure from the linear geostrophic balance on the horizontal plane. 
That the truncated QG systems satisfy the Liouville theorem, thus 
allowing the use of Gibbs ensemble methods, can be seen, e.g., in 
Ref.~\cite{Majda2006}.

A particular case of the QG models is the Surface Quasi-Geostrophic (SQG) 
approximation 
\cite{Pierrehumbert1994,Held1995,Tran2004,Tran2005,Tran2006a,Tran2006b}, 
which describes rotating stratified flows with constant potential vorticity. 
In this model, the vertical gradient of the stream function $\psi$ matches 
a scalar field (the density field $\rho$ or the potential temperature $T$) 
at a flat surface $z=0$. The scalar field is identified with the 
horizontal Laplacian $(-\triangle)^{1/2} \psi$, and the equation for 
the advection of this scalar by the incompressible surface flow 
${\bf u} = \hat{z} \times \nabla \psi   = (-\partial_y \psi,\partial_x \psi,0)$ 
is solved. Contrary to 3D QG turbulence whose dynamics is driven by 
large-scale interior potential vorticity gradients, SQG flows are entirely 
driven by density or potential temperature variations at the surface. 
Recently, this system has been used to study the dynamics of the upper 
troposphere \cite{Juckes1994,Hakim2002,Tulloch2006}, and of the upper 
oceanic layers with relative accuracy down to $500$ meters 
\cite{Lapeyre2006,LaCasce2006,Isern2006}. Turbulence in forced-dissipative 
SQG systems (including direction of cascades and scaling laws) has also 
been studied in numerical simulations 
\cite{Pierrehumbert1994,Tran2005,Tran2006b}.

Besides potential applications in ocean and atmospheric dynamics, 
the SQG equations have interest from the mathematical and physical points 
of view. Although the SQG equations are 2D in nature, their dynamics 
share similarities with the 3D Euler equations, and whether a singularity 
in the gradients of the scalar field develops at finite time from smooth 
initial conditions has been the focus of many mathematical and numerical 
studies \cite{Constantin1994,Constantin1999,Constantin2001,Cordoba2004}. 
Although there is no clear-cut answer for the most general case, a 
singularity has been ruled out for the case in which isolevels of the 
scalar contain a hyperbolic saddle \cite{Cordoba1998}. In the same 
context, numerical comparisons between the dynamics of the inviscid and 
viscous SQG equations were carried in \cite{Ohkitani1997}. More recently, 
cascades in the SQG equations have received attention from the physical 
point of view, as it was found that large-scales in SQG flows have 
conformal invariant properties \cite{Bernard2007}.

In this paper we investigate the classical Gibbs ensemble solution of 
the truncated inviscid SQG equations, and compare these solutions and their 
viscous-like transient with solutions of the dissipative SQG equations. 
In particular, we derive solutions for the pseudo-energy and 
pseudo-enstrophy equilibrium spectrum, verify numerically the convergence 
of the truncated SQG solutions to the statistical equilibrium solutions, 
and  compare inviscid and viscous numerical results. The statistical 
equilibrium solution indicates the pseudo-enstrophy thermalizes while the
pseudo-energy condenses in SQG, in agreement with previous numerical 
results where a direct cascade of pseudo-enstrophy and an inverse cascade 
of pseudo-energy were reported \cite{Pierrehumbert1994,Smith2002}. In the 
viscous-like transient, we observe the development of inertial ranges with 
$\sim k^{-5/3}$ scaling for the spectrum of the pseudo-enstrophy, 
unlike previous inviscid simulations where steeper power laws were 
observed \cite{Ohkitani1997}. Finally, we present an analogy between the 
non-thermalized fraction of the pseudo-energy and pseudo-enstrophy in the 
inviscid truncated runs, and the free decay of the pseudo-energy and 
pseudo-enstrophy in the viscous SQG equations at very large Reynolds 
number. The analogy allows us to identify possible asymptotic 
behavior for very large Reynolds numbers.

The paper is organized as follows. In Sec.~\ref{sec:Theory} we introduce 
the SQG equations and derive the statistical equilibrium solutions using the 
canonical distribution function formalism. In Sec.~\ref{sec:NumRes} we 
present a set of numerical simulations of the truncated inviscid SQG 
equations. We first compare them with the theoretical equilibrium solutions, 
and we then introduce a viscous SQG system comparing the time evolution 
of viscous and inviscid simulations. Finally, we present our conclusions in 
Sec.~\ref{sec:Conclusions}. 

\section{\label{sec:Theory}SQG and the Gibbs ensemble}

The system considered in this paper is an incompressible 2D SQG flow. 
Its equations are usually presented in the literature as part of a family 
of equations governing the advection of a scalar 
\cite{Pierrehumbert1994}
\begin{equation}
q = (-\triangle)^{\alpha/2}\psi.
\label{eq:cero}
\end{equation}
This family includes the SQG equations 
($\alpha=1$, which is the case in the present study) 
\cite{Blumen1978,Held1995}, the vorticity equation in an Euler flow 
($\alpha=2$), and the equation of motion for a shallow flow in a rotating 
domain driven by uniform internal heating ($\alpha=3$) \cite{Tran2004}. 
The case $\alpha = - 2$ corresponds 
to a shallow-water QG equation in the 
limit of large length scales compared to the deformation 
scale (see \cite{Larichev1991}). 

The flow in these models is described by a stream function $\psi$, 
and governed by the so-called $\alpha$-turbulence equations which, in their 
inviscid unforced version, are usually written in the general form
\begin{equation}
\partial_t q +J(\psi,q)=0,
\label{eq:primera}
\end{equation}
where $J$ is the Poisson bracket
\begin{equation}
J(A,B)=\partial _x A\partial _y B -\partial  _x B \partial  _y A.
\label{eq:segunda}
\end{equation}
For SQG, Eq.~(\ref{eq:cero}) in Fourier space reduces to 
$\hat{q}({\bf k})=|{\bf k}|\hat{\psi}({\bf k})$, and 
Eq.~(\ref{eq:primera}) reads
\begin{equation}
\partial_t \hat{\psi} = \frac{1}{|{\bf k}|}
    \left(\widehat{\partial_xq\, \partial_y\psi}-
    \widehat{\partial_x\psi\, \partial_yq}\right) ,
\label{eq:ec_mov}
\end{equation}
where the hats denote Fourier transformed. Inviscid SQG dynamics 
possesses two quadratic conserved quantities, analogous to the energy 
and enstrophy in 2D Euler flows, which are defined respectively as
\begin{equation}
E = - \frac{1}{A}\int q \psi \, dx dy,
\end{equation}
and
\begin{equation}
G = \frac{1}{A}\int q^2 \, dx dy,
\end{equation}
with $A$ the total area of the integration domain. We will refer to 
these magnitudes as the pseudo-energy and pseudo-enstrophy respectively. 
Note that sometimes these quantities are defined with a factor 
$1/2$ in front, in analogy with the energy and enstrophy in the 2D Euler 
equations (see, e.g., \cite{Blumen1978}). For the sake of simplicity, 
in the following we use the convention in \cite{Pierrehumbert1994,Held1995} 
and drop those factors.

Note in the forced-dissipative case, Kolmogorov-Batchelor-Kraichnan 
phenomenology predicts two inertial ranges for these quantities 
\cite{Pierrehumbert1994}: an inverse cascade of pseudo-energy with spectra 
$E(k)\sim k^{-2}$ and $G(k) \sim k^{-1}$, and a direct cascade of 
pseudo-enstrophy with spectra $E \sim k^{-8/3}$ and $G \sim k^{-5/3}$. 
We will consider this case in Sec.~\ref{sec:NumRes}.

Writing $k = |{\bf k}|$, the pseudo-energy and pseudo-enstrophy of each 
Fourier mode ${\bf k}$ are easily related by
\begin{equation}
G({\bf k})=kE({\bf k})=|\hat{u}({\bf k})|^2.
\end{equation}
Since $\hat{u}_x({\bf k})=-ik_y\hat{\psi}({\bf k})$, 
$\hat{u}_y({\bf k})=ik_x\hat{\psi}({\bf k})$, and the relation between the 
pseudo-energy in each mode and the stream function is
\begin{equation}
|\hat{u}|^2=(k^2_x+k^2_y)|\hat{\psi}|^2=k^2|\hat{\psi}|^2,
\label{eq:uphi}
\end{equation}
the pseudo-energy and pseudo-enstrophy can then be written in Fourier 
space as follows
\begin{equation}
E=\sum_{{\bf k}/k \epsilon N'} k|\hat{\psi}({\bf k})|^2,
\label{eq:Et}
\end{equation}
\begin{equation}
G=\sum_{{\bf k}/k \epsilon N'} k^2|\hat{\psi}({\bf k})|^2,
\label{eq:Gt}
\end{equation}
where $N' = \{1,...,N/2\}$ runs over all degrees of freedom of the system. 
Note that in Eqs.~(\ref{eq:Et}) and (\ref{eq:Gt}) we already truncated the 
system up to a finite number of modes.

From these relations we can derive the statistical equilibrium solutions 
for the inviscid truncated SQG system. The generalized Gibbs canonical 
distribution is
\begin{equation}
P=\frac{1}{Z}e^{-(\beta E + \gamma G)},
\end{equation}
where
\begin{equation}
Z=\int_{\textrm{phase space}} e^{-(\beta E + \gamma G)} d\xi,
\end{equation}
is the partition function and 
$d\xi = \prod_{\bf k} d{\bf u}_1({\bf k})d{\bf u}_2({\bf k})$; ${\bf u}_1$ and 
${\bf u}_2$ are defined such that 
$\hat{{\bf u}}({\bf k})={\bf u}_1({\bf k})+i{\bf u}_2({\bf k})$ and are 
constrained by the incompressibility condition $\nabla \cdot {\bf u} = 0$ 
(or ${\bf k} \cdot \hat{{\bf u}}=0$ in Fourier space) by
\begin{equation}
k_x\hat{u}_{1x} + k_y\hat{u}_{1y} = 0,
\end{equation}
\begin{equation}
k_x\hat{u}_{2x} + k_y\hat{u}_{2y} = 0.
\end{equation}

From Eqs.~(\ref{eq:uphi}), (\ref{eq:Et}), and (\ref{eq:Gt}),
\begin{equation}
E = \sum_{{\bf k}/k \epsilon N'}\frac{|\hat{{\bf u}}({\bf k})|^2}{k},
\end{equation}
\begin{equation}
G = \sum_{{\bf k}/k \epsilon N'}|\hat{{\bf u}}({\bf k})|^2,
\end{equation}
and the partition function results
\begin{eqnarray}
Z  &=& \int e^{-\beta\sum\frac{|{\bf u}({\bf k})|^2}{k}
    -\gamma \sum|{\bf u}({\bf k})|^2}\prod_{{\bf k}/k \epsilon N'} 
d{\bf u}_1({\bf k})d{\bf u}_2({\bf k}) 
    \nonumber \\
{} &=& \prod_{{\bf k}/k \epsilon N'} 2\pi \int 
    e^{-|{\bf u_1}({\bf k})|^2(\frac{\beta}{k}+\gamma)}|{\bf u_1}({\bf k})| 
    du_1({\bf k}) \times \nonumber \\
{} &{}& \times \,\, 2\pi \int e^{-|{\bf u_2}({\bf k})|^2
    (\frac{\beta}{k}+\gamma)}|{\bf u_2}({\bf k})|du_2({\bf k}) .
\end{eqnarray}
Note we used isotropy of the velocities ${\bf u_1}$ and ${\bf u_2}$ 
to convert the first integral in the entire phase space into two 
one-dimensional integrals with respect to the absolute values 
$u_1=|{\bf u_1}|$ and $u_2=|{\bf u_2}|$ respectively.

Using 
\begin{equation}
\int_0^{\infty} e^{-x^2(\frac{\beta}{k}+\gamma)}xdx=
    \frac{k}{2(\gamma k + \beta)},
\label{eq:integral}
\end{equation}
which holds provided $Re(\gamma+\beta/k)>0$, the partition function 
results
\begin{equation}
Z = \prod_{{\bf k}/k \epsilon N'}\left( \frac{\pi k}{\gamma k + \beta} 
\right)^2 = \prod_{{\bf k}/k \epsilon N'} Z_{\bf k}
\end{equation}
The most probable probability density function can be obtained by 
other methods. A Lagrange multiplier method can be used (see 
\cite{Majda2006}). Another elegant way is to use properties of 
$n$-dimensional normal distributions (see, e.g., \cite{Servidio2008}). 
All these methods yield the same result.

From the partition function we can derive the  mean 
modal intensity spectra of pseudo-energy and pseudo-enstrophy
\begin{equation}
E({\bf k}) = -\frac{\partial \ln Z_{\bf k}}{\partial \beta}=
    \frac{2}{\gamma k +\beta},
\label{eq:Emod}
\end{equation}
\begin{equation}
G({\bf k}) =-\frac{\partial \ln Z_{\bf k}}{\partial \gamma}=
    \frac{2k}{\gamma k + \beta}.
\label{eq:Gmod}
\end{equation}
These spectra give the pseudo-energy and pseudo-enstrophy in each mode 
${\bf k}$, and are functions of its modulus only. The usual isotropic 
spectrum is obtained by integrating Eqs.~(\ref{eq:Emod}) and (\ref{eq:Gmod}), 
resulting, e.g., $E(k) = \pi k E({\bf k})$ \cite{Kraichnan1975} provided the 
modes are dense enough over the entire spectrum. We then finally get the 
expressions for the isotropic spectra of pseudo-energy and pseudo-enstrophy,
\begin{equation}
E(k)=\int E({\bf k}) k d\varphi = 
    \pi k E({\bf k}) = \frac{2\pi k}{\gamma k +\beta},
\label{eq:Eiso}
\end{equation}
\begin{equation}
G(k)=\int G({\bf k}) k d\varphi = 
    \pi k G({\bf k}) = \frac{2\pi k^2}{\gamma k +\beta}.
\label{eq:Giso}
\end{equation}

The pseudo-energy and pseudo-enstrophy are quadratic magnitudes and therefore 
Eqs.~(\ref{eq:Eiso}) and (\ref{eq:Giso}) must be positive. The relation 
$\gamma k > -\beta$ must apply to every value of $k$, and as the case $k=1$ 
is the more restrictive, we simply need the system to satisfy the condition
\begin{equation}
\gamma > - \beta.
\label{eq:condition}
\end{equation}
This condition is enough for integral (\ref{eq:integral}) to converge. 
Furthermore, asking the total pseudo-energy and pseudo-enstrophy to be 
positive, it is obtained that $\gamma>0$ and that the pseudo-enstrophy 
thermalizes.

A truncated inviscid SQG system is expected to reach at large times the 
absolute equilibrium solutions (\ref{eq:Eiso}) and (\ref{eq:Giso}). The 
values of $\gamma$ and $\beta$ are uniquely determined by the total amount 
of pseudo-energy and pseudo-enstrophy contained in the initial conditions, 
and can be calculated solving the system of equations
\begin{equation}
E(t=0) = \sum_k\frac{2\pi k}{\gamma k + \beta} ,
\label{eq:Etotal}
\end{equation}
\begin{equation}
G(t=0) = \sum_k\frac{2\pi k^2}{\gamma k + \beta} .
\label{eq:Gtotal}
\end{equation}
Note these equations can be easily converted to two polynomial 
equations with two unknowns ($\gamma$ and $\beta$). The number of terms 
on the r.h.s.~of Eqs. (\ref{eq:Etotal}) and (\ref{eq:Gtotal}) depend 
on the maximum wavenumber preserved after the truncation (and therefore, 
in numerical simulations, on the linear resolution $N$). As $E(t=0)$ 
and $G(t=0)$ are known, the system can be easily solved numerically. 
For the resolutions used in the next section, the values obtained for 
$(\gamma , \beta)$ are: 
$(346.23, -342.60)$ for $N=32$, 
$(1344.79, -1341.39)$ for $N=64$, 
$(5044.05, -5040.75)$ for $N=128$, 
$(19504.20,-19500.90)$ for $N=256$, 
$(77579.35,-77576.13)$ for $N=512$, and
$(311246.27,-311243.06)$ for $N=1024$.

\begin{figure}
\includegraphics[width=9cm]{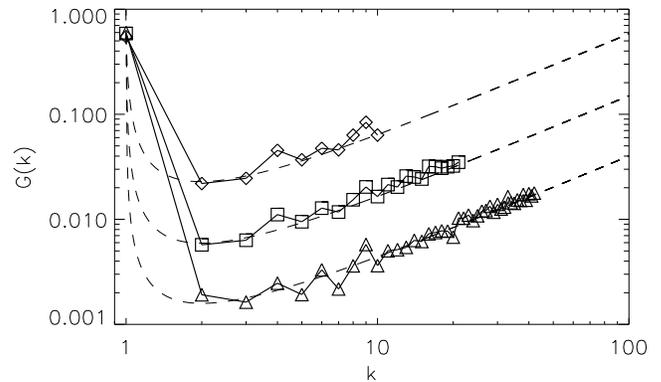}
\caption{Isotropic pseudo-enstrophy spectra for runs with $N=32$ (diamonds), 
64 (squares), and $128$ (triangles) at $t=20$, 1000, and 24000 
respectively. The Gibbs ensemble prediction (\ref{eq:Giso}) for each run is 
shown in dashed lines. The values of $(\gamma, \beta)$ obtained from 
solving Eqs.~(\ref{eq:Etotal}) and (\ref{eq:Etotal}) for the initial 
pseudo-energy and pseudo-enstrophy of the runs are $(346.23, -342.60)$ for 
$N=32$, $(1344.79, -1341.39)$ for $N=64$, and $(5044.05, -5040.75)$ for 
$N=128$. Theoretical and numerical results are in good agreement for all 
wavenumbers.}
\label{fig:fiteos1}
\end{figure}

\begin{figure}
\includegraphics[width=9cm]{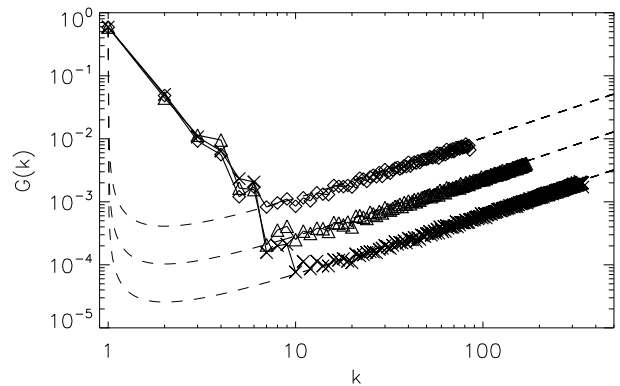}
\caption{Isotropic pseudo-enstrophy spectra at $t=200$ for runs with 
$N=256$ (diamonds), 512 (triangles), and 1024 (crosses). The Gibbs ensemble 
prediction (\ref{eq:Giso}) is shown in dashed lines. The values of 
$(\gamma, \beta)$ obtained from solving Eqs.~(\ref{eq:Etotal}) and 
(\ref{eq:Etotal}) for the initial pseudo-energy and pseudo-enstrophy of 
the runs are $(19504.20,-19500.90)$ for $N=256$, $(77579.35,-77576.13)$ 
for $N=512$, and $(311246.27,-311243.06)$ for $N=1024$. As the resolution 
increases, it takes longer times for the equilibrium at large scales to be 
reached, while thermalization at large wave numbers is achieved fast.}
\label{fig:fiteos2}
\end{figure}

\section{\label{sec:NumRes}Numerical Results}

In this section we present numerical simulations of the SQG equations. We 
first compare the inviscid spectra at late times stemming from the 
simulations with the results derived above. Then, we study the transition 
from the initial condition towards the equilibrium spectrum in simulations 
at larger spatial resolution, and compare the behavior of the non-thermalized 
fraction of pseudo-energy and pseudo-enstrophy in the inviscid runs with the 
pseudo-energy an pseudo-enstrophy in viscous runs.

\begin{figure}
\includegraphics[width=8cm]{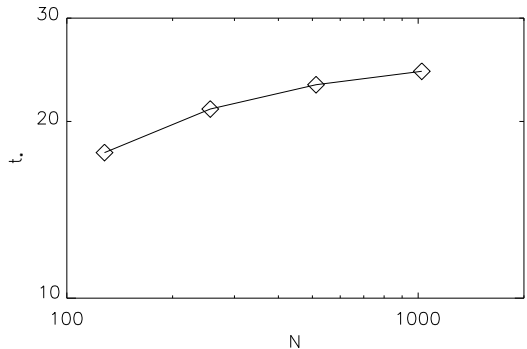}
\caption{Time $t_*$ for $50\%$ of the pseudo-enstrophy to thermalize in 
the inviscid runs as a function of the resolution $N$.}
\label{fig:tstar} 
\end{figure}

\subsection{Inviscid truncated runs}

To solve numerically Eq.~(\ref{eq:ec_mov}) in a 2D periodic domain of 
length $2\pi$, we use a parallel pseudospectral code 
\cite{Gomez2005,Mininni2011b}. Second order Runge-Kutta is used to evolve 
in time, and non-linear terms are computed with the $2/3$-rule for 
dealiasing, so that the Fourier space is truncated at the maximum wave 
number $k_{max}=N/3$, with $N$ the linear resolution. Under these 
conditions, pseudospectral methods are known to be equivalent to Galerkin 
spectral methods \cite{Canuto2006}, and all quadratic invariants of the 
original equations are conserved in the truncated Fourier space. As an 
example, in simulations of the inviscid SQG equations with resolution of 
$512^2$ grid points and a time step of $\Delta t = 2\times 10^{-4}$, the 
pseudo-energy was conserved up to the fourth digit after $500$ turnover 
times. 

Spatial resolutions ranged from $32^2$ to $1024^2$ grid points. The initial 
condition for all runs is a superposition of Fourier modes with random phases, 
with a pseudo-energy spectrum $E(k) \sim k^{-2}$ for $1 \le k \le 4$, and 
zero otherwise. All simulations behave in a similar fashion, showing a 
progressive thermalization of modes with higher wavenumber and reaching 
equilibrium for long times; however, simulations at larger resolution take 
longer time to reach equilibrium at all scales. Once the system reaches 
equilibrium, its spectra should be compatible with solutions (\ref{eq:Eiso}) 
and (\ref{eq:Giso}). To compare simulation results  with the theoretical 
predictions, $\gamma$ and $\beta$ were determined solving the set of 
Eqs.~(\ref{eq:Etotal}) and (\ref{eq:Gtotal}) for each spatial resolution 
separately, as explained in the previous section.

\begin{figure}
\includegraphics[width=8cm]{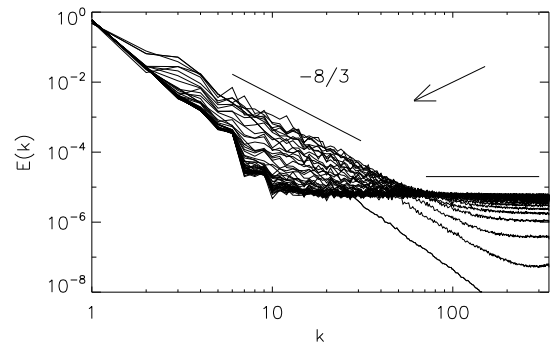}
\caption{Evolution of the pseudo-energy spectra for the inviscid run with 
$N=1024$ from $t=1$ to $t=205$ with time increments $\Delta t = 4$. A 
$k^{-8/3}$ power law followed by an equilibrium subrange is observed for 
early times. As time evolves, more modes approach the equilibrium spectrum. 
The Gibbs ensemble prediction is indicated as a reference by the horizontal 
line. The arrow indicates the direction in which the spectrum evolves in 
time.}
\label{fig:E_tmedios} 
\end{figure}

\begin{figure}
\includegraphics[width=8cm]{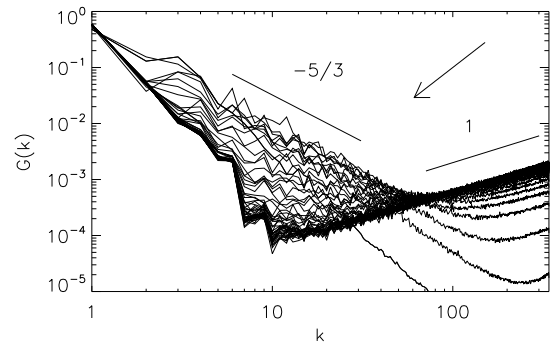}
\caption{Evolution of the pseudo-enstrophy spectra for the inviscid run with 
$N=1024$ from $t=1$ to $t=205$ with time increments $\Delta t = 4$. A 
$k^{-5/3}$ power law followed by a thermalized subrange is observed for 
early times. As time evolves, more modes achieve thermalization. The 
Gibbs ensemble prediction is indicated as a reference by the $\sim k$ line. 
The arrow indicates the direction in which the spectrum evolves in time.}
\label{fig:G_tmedios} 
\end{figure}

Numerical results for the isotropic pseudo-enstrophy spectrum once the 
equilibrium was reached are shown in Fig.~\ref{fig:fiteos1}, with different 
symbols for the numerical data, and with dashed lines for the theoretical 
predictions. Numerical spectra are in good agreement with the theory at all 
wavenumbers, and the spectrum shows a peak at $k=1$ (associated with the 
condensation of pseudo-energy at the gravest modes), and a $\sim k$ scaling 
(associated with the thermalization of pseudo-enstrophy) for larger 
wavenumbers. However, note that as the resolution is increased, the time 
to reach the equilibrium increases.

Results for higher resolutions (from $256^2$ to $1024^2$ grid points) at 
$t=200$ together with their theoretical predictions are shown separately in 
Fig.~\ref{fig:fiteos2}. For higher resolutions, the thermalization process 
is slower and convergence of lower wavenumber modes to the statistical 
equilibrium solution takes progressively longer times. Nevertheless, 
numerical and theoretical results agree well for intermediate to high 
wavenumbers.

The slow down of the thermalization as resolution is increased can 
also be seen in Fig.~\ref{fig:tstar}, that shows the time $t_*$ for the 
system to have $50\%$ of its pseudo-enstrophy thermalized. Note this 
thermalization takes place predominantly at large wavenumbers, and to 
reach the equilibrium at lower wavenumbers takes substantially longer 
times.

Simulations at lower resolution provide as a result a faster way to verify 
the validity of the Gibbs ensemble prediction. However, to study the 
viscous-like transient that develops as the system evolves towards 
equilibrium, the runs with larger resolutions will allow us better 
identification of scaling laws and comparison with viscous runs. We 
therefore focus in the following on the $1024^2$ run. Figures 
\ref{fig:E_tmedios} and \ref{fig:G_tmedios} show the time evolution of 
the pseudo-energy and pseudo-enstrophy spectra in this run. At early times 
both spectra develop a viscous-like inertial range, with slopes compatible 
with Kolmogorov-Batchelor-Kraichnan phenomenology \cite{Pierrehumbert1994}. 
A power law $\sim k^{-5/3}$ can be identified in the pseudo-enstrophy, and 
$\sim k^{-8/3}$ in the pseudo-energy spectrum. As time evolves, the 
pseudo-enstrophy shows a progressive thermalization starting from the 
largest wavenumbers, and the viscous-like inertial range becomes narrower 
as the $G(k) \sim k$ thermalized spectrum broadens. We will take the 
minimum of the pseudo-enstrophy spectrum, at $k=k_{th}$, as the delimiting 
wavenumber between these two subranges. The flat pseudo-energy spectrum 
for $k>k_{th}$, and its peak at $k=1$, evidences a condensation of this 
quantity at low $k$ rather than a thermalization at high wavenumbers. 

\begin{figure}
\includegraphics[width=9cm]{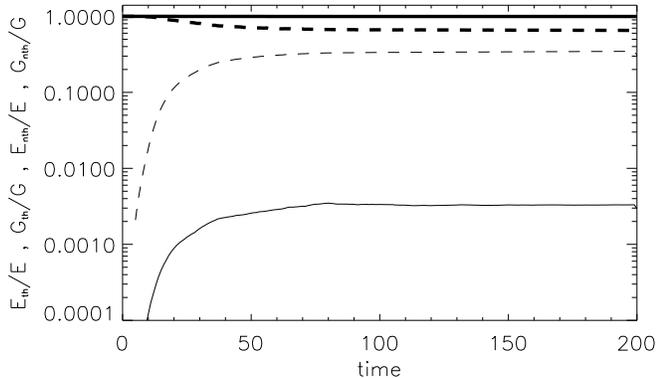}
\caption{Time evolution of the thermalized pseudo-energy (thin solid) 
and pseudo-enstrophy (thin dashed line) normalized by the total 
pseudo-energy and pseudo-enstrophy for the $1024^2$ run. Both magnitudes 
grow monotonically towards their equilibrium asymptotic value. 
Non-thermalized pseudo-energy (thick solid) and pseudo-enstrophy (thick 
dashed) normalized respectively by total pseudo-energy and 
pseudo-enstrophy are also shown.}
\label{fig:EthGth1024} 
\end{figure}

\begin{figure}
\includegraphics[width=9cm]{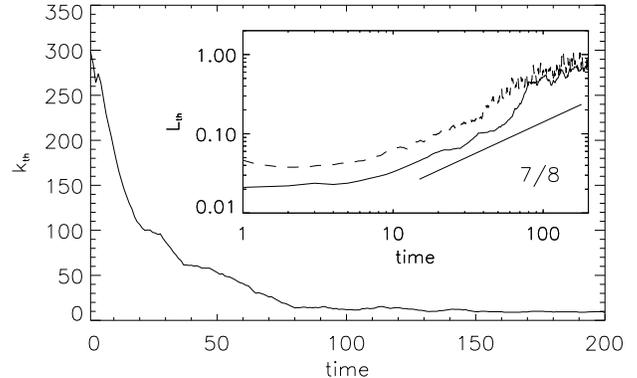}
\caption{Time evolution of $k_{th}$ for the inviscid run with $N=1024$. 
As time evolves, more modes reach thermalization, but $k_{th}$ does not 
converge to zero as condensation of pseudo-energy takes place at the 
gravest modes. Inset: time evolution of $L_{th}=2 \pi/k_{th}$ for the 
inviscid runs with $N=1024$ (solid) and with $N=512$ (dashed) with a 
$\sim t^{7/8}$ slope shown as a reference.}
\label{fig:kth1024} 
\end{figure}

In the following we define the thermalized pseudo-energy and 
pseudo-enstrophy respectively as the sum of the pseudo-energy and 
pseudo-enstrophy contained in all modes with $k \ge k_{th}$,
\begin{equation}
E_{th}= \sum_{k=k_{th}}^{k_{max}}E(k), \,\,\,\, 
    G_{th}=\sum_{k=k_{th}}^{k_{max}}G(k).
\label{eq:Eth}
\end{equation}
In a similar way, we define the non-thermalized pseudo-energy and 
pseudo-enstrophy as the difference between the total amount of these 
quantities (which is constant in time) and their thermalized values,
\begin{equation}
E_{nth}=E-E_{th}, \,\,\,\,
    G_{nth}=G-G_{th}.
\label{eq:nthEG}
\end{equation}

The time evolution of $E_{th}$, $E_{nth}$, $G_{th}$, and $G_{nth}$ normalized 
respectively by $E$ and $G$ is shown in Fig.~\ref{fig:EthGth1024}. $E_{th}$ 
and $G_{th}$ grow monotonically at early times, converging to an almost 
constant value for long times. This is consistent with 
Figs.~\ref{fig:E_tmedios} and \ref{fig:G_tmedios}, where more and more 
modes approach the statistical equilibrium solution as time evolves. This 
is also illustrated by the evolution of $k_{th}$ in Fig.~\ref{fig:kth1024}. 
As most of the pseudo-energy condenses at $k=1$, $E_{th}$ remains a small 
fraction of the total pseudo-energy while $E_{nth}$ (which in this case 
represents the condensed pseudo-energy) stays almost constant. On the 
other hand, $G_{th}$ grows to a larger fraction of the total 
pseudo-enstrophy as the $\sim k$ thermalized spectrum for $G$ at large 
wavenumbers (see Fig.~\ref{fig:G_tmedios}) holds a significant 
fraction of the pseudo-enstrophy in the system.

As shown in Ref.~\cite{Krstulovic2008} for the truncated Euler equations, 
although the truncated system is inviscid, the transfer of 
pseudo-enstrophy towards thermalized modes before the equilibrium is 
reached (resulting in the growth of $E_{th}$ and $G_{th}$ in time) can be 
interpreted as a viscous-like transient in which the thermalized modes 
give rise to an effective viscosity acting on the non-thermalized range. 
This effective viscosity is responsible for the development of turbulent 
inertial subranges. In the next subsection we analyze in more detail 
this transient, comparing the inviscid system with viscous solutions.

\begin{figure}
\includegraphics[width=8cm]{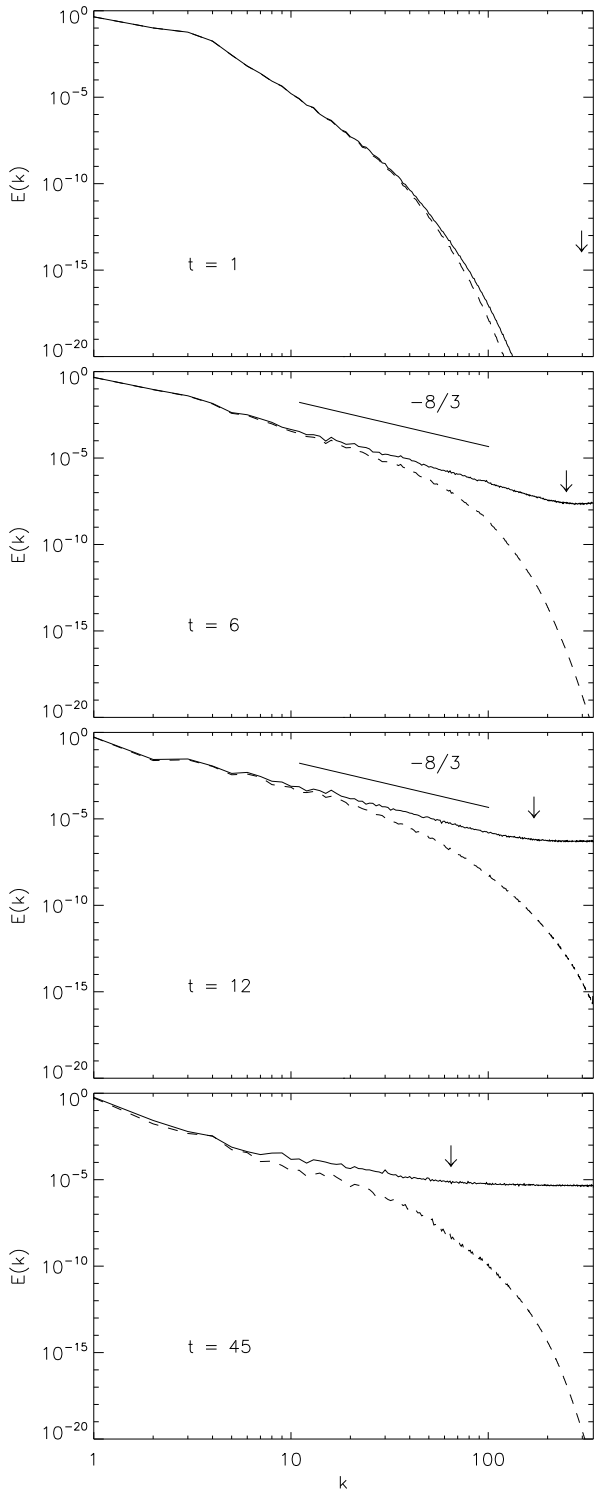}
\caption{Pseudo-energy spectra at different times for $1024^3$ inviscid 
(solid) and viscous (dashed) runs. Both runs develop a power law 
compatible with $\sim k^{-8/3}$ (indicated as a reference by the 
straight line). Later on, the spectra differ at intermediate and large 
wavenumbers, with the gravest modes still showing similar amplitudes. 
The arrows indicate the values of $k_{th}$ at each time.}
\label{fig:espectros1024} 
\end{figure}

\begin{figure}
\includegraphics[width=8cm]{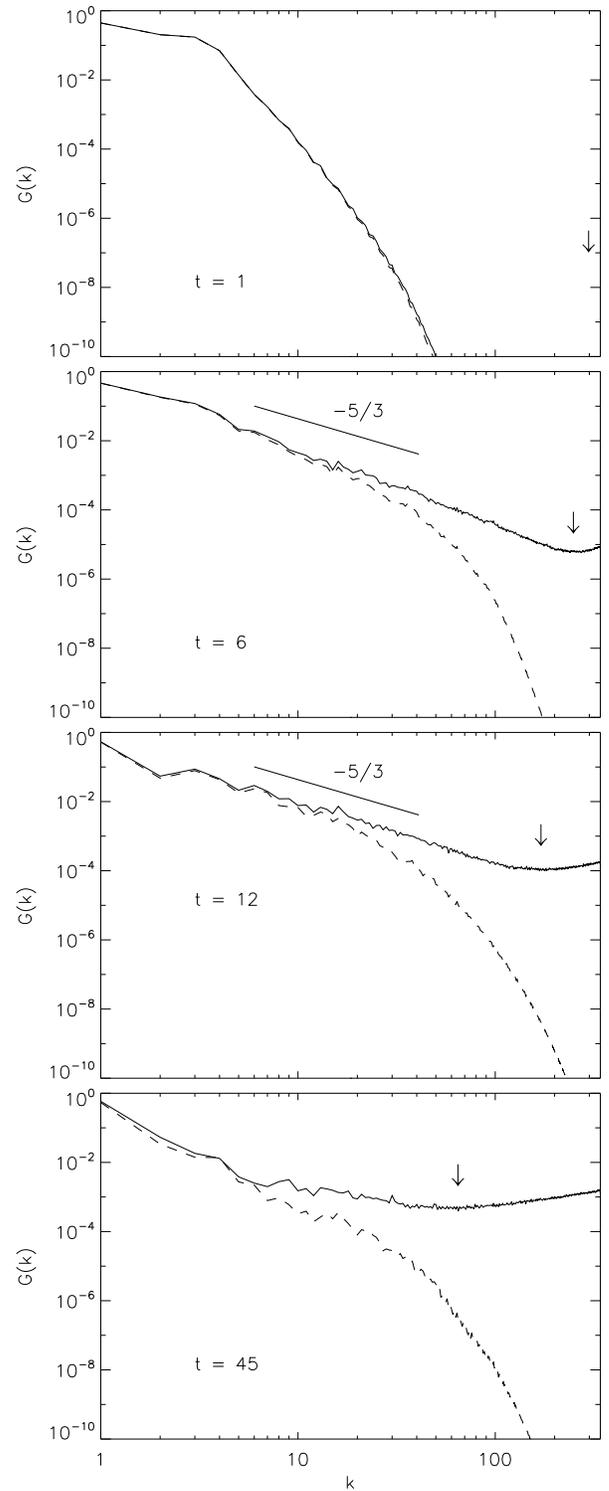}
\caption{Pseudo-enstrophy spectra at different times for the same $1024^3$ 
inviscid (solid) and viscous (dashed) runs shown in 
Fig.~\ref{fig:espectros1024}. A $\sim k^{-5/3}$ slope is shown as a 
reference. The arrows indicate the values of $k_{th}$ at each time.}
\label{fig:espectros1024G} 
\end{figure}

\subsection{Inviscid vs. dissipative systems}

We focus now on the period of time when $E_{th}$ and $G_{th}$ are time 
dependent, and the flux of pseudo-enstrophy towards thermalized modes can be 
interpreted as an out-of-equilibrium turbulent solution for $k<k_{th}$, 
with effective viscosity associated with the thermalized modes with 
$k>k_{th}$. We show that during this period, the ideal truncated model 
can give valuable information on the behavior of a similar but 
dissipative system (by similar, we mean a viscous system subject to the 
same initial conditions). With this aim, we compare ideal and 
dissipative spectra and decays for this particular period of time. 

To consider dissipative SQG flows, we must solve Eq.~(\ref{eq:ec_mov}) 
with the addition of a dissipative term,
\begin{equation}
\frac{\partial \hat{\psi}}{\partial t} = \frac{1}{|{\bf k}|}
    \left(\widehat{\partial_xq \, \partial_y\psi}-
    \widehat{\partial_x\psi \, \partial_yq} \right)- 
    \nu|{\bf k}|^2\hat{\psi}.
\label{eq:ec_mov_dis}
\end{equation}
We solve this equation in the same way as in the ideal case, with a 
pseudospectral method with the $2/3$-rule for dealiasing, and with second 
order Runge-Kutta to evolve in time.

Equation (\ref{eq:ec_mov_dis}) in terms of the scalar $q$ in real space 
reads 
\begin{equation}
\frac{\partial q}{\partial t} = -{\bf u}\cdot \nabla q + \nu \nabla^2 q
\label{eq:ec_q}
\end{equation}
Multiplying the equation by $q$ the following conservation law follows
\begin{equation}
\frac{1}{2}\frac{\partial q^2}{\partial t} = 
    -\frac{1}{2}\nabla \cdot (q^2 {\bf u}) + \nu q\nabla^2 q.
\label{eq:ec_q2}
\end{equation}
Integrating over the entire domain, the first term on the r.h.s. cancels and 
Eq.~(\ref{eq:ec_q2}) in Fourier space results
\begin{equation}
\frac{dG}{dt}=\frac{d}{dt} \int_{0}^{\infty} G(k)dk = 
    - 2\nu \int_{0}^{\infty} k^2 G(k) dk = -\sigma
\label{eq:ec_q3}
\end{equation}
with $\sigma$ the pseudo-enstrophy dissipation rate. Introducing a 
dissipation wavenumber $k_{\eta}$ such that all dissipation is 
concentrated in Fourier space between $k=1$ and $k=k_{\eta}$, in the 
turbulent steady state we can approximate
\begin{equation}
\sigma \approx 2\nu \int_{1}^{k_{\eta}}k^2G(k) dk.
\label{eq:gamma}
\end{equation}
From Ref.~\cite{Pierrehumbert1994}, $G(k) \sim \sigma^{2/3}k^{-5/3}$,  
and replacing in Eq.~(\ref{eq:gamma}) we finally get
\begin{equation}
k_{\eta} \sim \left(\frac{\sigma}{\nu^3}\right)^{1/4}.
\end{equation}
This relation is important to fix the resolution in viscous simulations 
so that all relevant scales in the flow are well resolved. In practice, 
we want the dissipation wavenumber to be smaller than the maximum 
resolved wavenumber $k_{max}$, or in other words, we ask for the 
condition 
\begin{equation}
\left[\sum_{1}^{k_{max}}\frac{k^2 G(k)}{\nu^2}\right]^{1/4} \le \frac{N}{3}
\end{equation}
to be fulfilled at all times.

\begin{figure}
\includegraphics[width=7.7cm]{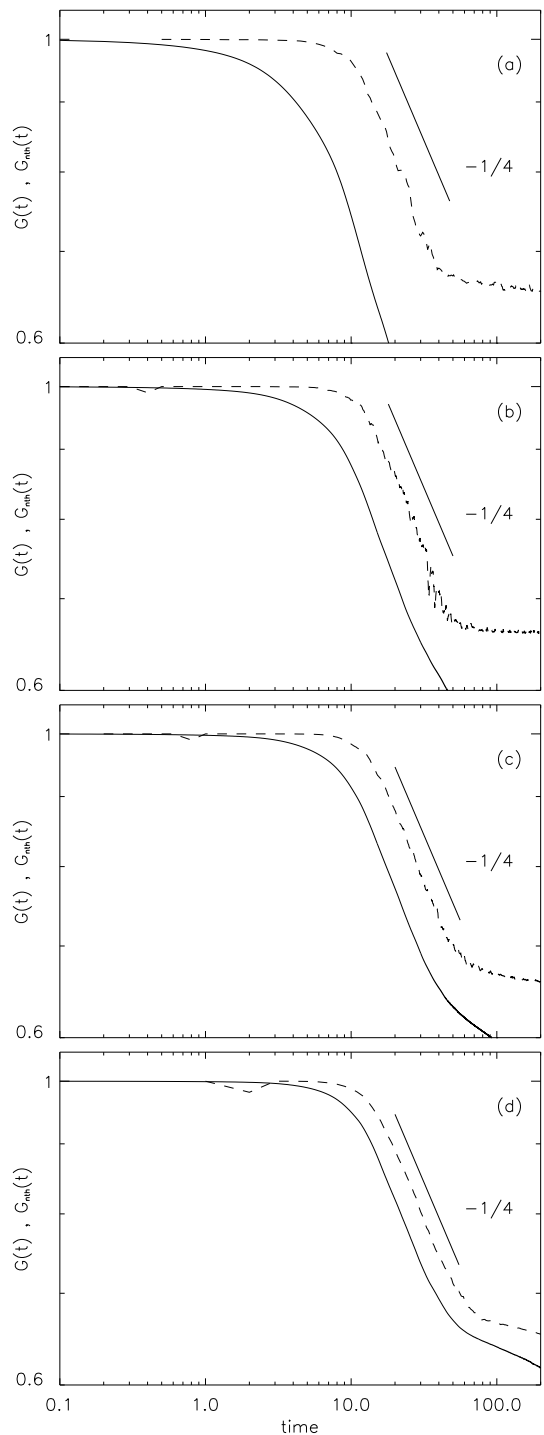}
\caption{Decay of pseudo-enstrophy $G(t)$ in viscous runs (solid line) 
and time evolution of $G_{nth}(t)=G-G_{th}(t)$ in ideal truncated runs 
(dashed lines) at increasing resolution: (a) $N=128$, $\nu=2 \times 10^{-3}$, 
(b) $N=256$, $\nu=5 \times 10^{-4}$, (c) $N=512$, $\nu=2.5 \times 10^{-4}$, 
and (d) $N=1024$, $\nu=9 \times 10^{-5}$.  The viscous $G(t)$ approaches 
$G_{nth}(t)$ as the Reynolds number is increased. A $\sim t^{-1/4}$ decay 
is indicated as a reference. The dissipation wavenumbers for each 
viscous run are: $k_\eta=36$ for $N=128$, $k_\eta=84$ for $N=256$, 
$k_\eta=150$ for $N=512$, and $k_\eta=318$ for $N=1024$, while the maximum 
resolved wavenumber is $k_{max}=N/3$.}
\label{fig:decay_todos} 
\end{figure}

\begin{figure}
\includegraphics[width=9cm]{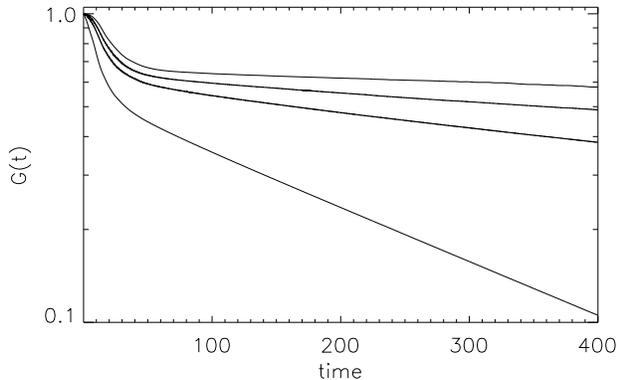}
\caption{Decay of pseudo-enstrophy $G(t)$ for viscous runs $N=1024$, $N=512$, 
$N=256$, $N=128$ from top to bottom in semi-log scale. Note the exponential 
decay after $t \approx 80$.}
\label{fig:Gexp}
\end{figure}

In Fig.~\ref{fig:espectros1024} we show the pseudo-energy spectra at 
different times for the $1024^2$ ideal truncated run, and for a $1024^2$ 
well resolved viscous run with $\nu=4\times 10^{-4}$. At early times, the 
pseudo-enstrophy contained at large scales is transferred to smaller scales, 
and is eventually affected by thermalization or by viscous dissipation 
in the ideal or viscous case respectively. These effects 
are different, as thermalization is a time dependent problem which 
results in a time dependent effective viscosity (see, e.g., 
\cite{Krstulovic2008,Krstulovic2009}), while viscosity in the dissipative 
run is fixed. However, the large scales of both systems are similar at 
low wavenumbers, with both systems developing a $\sim k^{-8/3}$ power law in 
the pseudo-energy in accordance with the theoretical results (although this 
subrange is wider at early times in the inviscid truncated run). 
This power law is lost at later times in the inviscid run, although the 
amplitude of the gravest modes still shows agreement with the 
viscous run. A similar behavior is observed in the time evolution 
of the pseudo-enstrophy spectrum (see Fig.~\ref{fig:espectros1024G}).

The similitude in the evolution of the large scale spectra in viscous and 
inviscid runs makes it plausible to compare the decay of quadratic 
quantities in the viscous simulations, say $G(t)$, with the evolution of the 
associated non-thermalized quantity $G_{nth}(t)$ in inviscid truncated runs. 
In freely decaying turbulent flows, quadratic quantities often develop a 
self-similar decay law in time, once turbulence develops and dissipation 
sets in. The power law obeyed during this decay (in this example, say, 
$G(t) \sim t^{-\Lambda}$) often requires large resolution as viscous effects 
tend to affect the decay making determination of the decay rate from 
the simulations difficult.

\begin{figure}
\includegraphics[width=9cm]{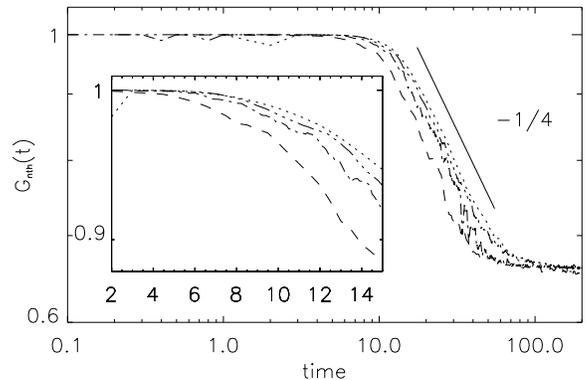}
\caption{Evolution of $G_{nth}$ for different grid sizes: $N=128$ 
(dashed), $256$ (dash-dotted), $512$ (dash-triple-dotted), and 
$1024$ (dotted). There is small delay in the time when the decay begins 
as the grid size (or equivalently, $k_{max}$) is increased, shown in more 
detail in the inset. Except for this delay, all runs show the same 
decay, with a $\sim t^{-1/4}$ decay shown as a reference.}
\label{fig:decay_Gth}
\end{figure}

In Fig.~\ref{fig:decay_todos}, we compare the decay of the pseudo-enstrophy 
in several dissipative runs with the evolution of $G_{nth}(t)$ in inviscid 
runs at different grid sizes. In all viscous cases, the pseudo-enstrophy is 
quasi-conserved at early times, and then a self-similar decay develops 
following a power law compatible (at the largest Reynolds numbers and 
spatial resolution) with $G(t) \sim t^{-1/4}$. It is worth noting 
that such a decay is also compatible with strict mathematical bounds for 
the decay of pseudo-enstrophy derived in \cite{Constantin1999,Cordoba2004}. 
The decay is followed by a saturation and a viscous exponential decay of 
the pseudo-enstrophy that remains in the system (see Fig.~\ref{fig:Gexp}).

The non-thermalized $G_{nth}(t)$ in all the inviscid runs shows the same 
evolution, which except for a small delay in the time when the 
quasi-conservation is broken (see Fig.~\ref{fig:decay_Gth} and inset), 
is independent of the spatial resolution considered. Remarkably, the 
evolution of $G(t)$ seems to approach, as the Reynolds number is 
increased, the behavior of $G_{nth}(t)$, to the point that the decay 
$\sim t^{-1/4}$ can be identified in $G_{nth}(t)$ at much lower resolution 
(e.g., grid sizes of $128^2$ or $256^2$ grid points) that in the viscous run.

\begin{figure}
\includegraphics[width=9cm]{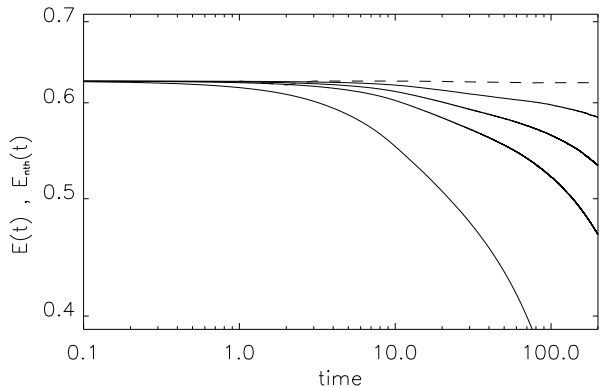}
\caption{Comparison between $E_{nth}$ in an $N=1024$ truncated inviscid 
simulation (dashed line) and $E(t)$ for viscous simulations with $N=128$ 
($\nu = 2 \times 10^{-3}$), $256$ ($\nu = 5 \times 10^{-4}$), $512$ 
($\nu = 2.5 \times 10^{-4}$), and $1024$ ($\nu = 9 \times 10^{-5}$) 
(bottom to top).}
\label{fig:decayE} 
\end{figure}

In this light, the decay of viscous SQG can be understood as follows. At 
early times, $G(t)$ remains approximately constant as the inertial range 
spectrum develops. As the smallest scales have not been excited yet, 
dissipative effects are negligible. Once turbulence develops, the decay 
law $G(t) \sim t^{-1/4}$ is observed. The decay rate in this stage is 
dominated by the turbulent cascade of the pseudo-enstrophy towards smaller 
scales, and the time evolution is given by the balance equation 
$dG/dt = -\sigma$ where the flux of pseudo-enstrophy $\sigma$ is controlled 
by the non-linear term in the equation, and independent of the value of 
the viscosity $\nu$ as long as the Reynolds number is large enough. As a 
result, the truncated inviscid run, although it has a different effective 
viscosity, shows the same decay. Later, all pseudo-enstrophy that remains 
in the system is associated with the condensation of pseudo-energy at 
$k=1$, and this remaining pseudo-energy (as well as the pseudo-enstrophy) 
can only decay exponentially in the viscous run (Fig.~\ref{fig:Gexp}), 
while it remains constant in the inviscid run (Fig.~\ref{fig:decay_Gth}).

Figure \ref{fig:decayE} compares the non-thermalized fraction of the 
pseudo-energy $E_{nth}(t)=E-E_{th}(t)$ in the $N=1024$ simulation with the 
decay of $E(t)$ in the viscous simulations with $N=128$, $256$, $512$, and 
$1024$. In the inviscid case, as most of the pseudo-energy condenses, 
there is no significant change in $E_{nth}$ with time. The viscous runs 
approach this behavior as the viscosity is decreased, in agreement with 
Batchelor-Kraichnan-Leith phenomenology for systems with condensation of 
one invariant at large scales (see, e.g., similar arguments for the 
decay of the enstrophy while the energy remains approximately constant 
in 2D Navier-Stokes \cite{Batchelor1969}). It is worth noting that 
a slow down in the decay of the pseudo-energy as the viscosity was 
decreased was already observed in \cite{Ohkitani1997}.

So far, random initial conditions were used in all runs. To analyze 
sensitivity of the decay to other initial conditions, we briefly discuss 
results for three initial conditions often used in studies of 
singularities in the QG equations 
\cite{Constantin1994,Ohkitani1997,Ohkitani2010}:
\begin{equation}
{\bf I)}\,\,\, q(t=0)=\sin (x) \sin (y) + \cos (y),
\label{eq:IC1}
\end{equation}
\begin{equation}
{\bf II)}\,\,\, q(t=0)=-(\cos (2x) \cos (y) + \sin (x) \sin (y)),
\label{eq:IC2}
\end{equation}
\begin{eqnarray}
{\bf III)}\,\,\, q(t=0) &=& \cos (2x) \cos (y)+ \sin (x) \sin (y)+
    \nonumber \\
    {} & {} & + \cos (2x) \sin(3y).
\label{eq:IC3}
\end{eqnarray}

\begin{figure}
\includegraphics[width=9cm]{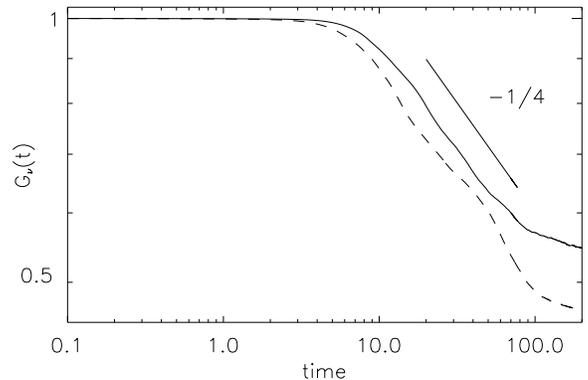}
\caption{Evolution of the pseudo-enstrophy for simulations with initial 
conditions ${\bf II}$ and ${\bf III}$, $N=1024$ and $\nu = 9 \times 10^{-5}$. 
Both runs show a decay compatible with $G(t)\sim t^{-1/4}$.}
\label{fig:decay_majda} 
\end{figure}

Initial conditions ${\bf II}$ and ${\bf III}$ lead to the same decay 
of the pseudo-enstrophy $\sim t^{-1/4}$ as observed in the previous 
simulations (see Fig.~\ref{fig:decay_majda}). However, initial condition 
${\bf I}$ results in a much slower decay (not shown), and up to the time 
we integrated the equations we were unable to identify any power law. 
Further analysis showed that the flow resulting from this initial 
condition, which initially activates only a few modes in Fourier space, 
develops a much steeper pseudo-energy and pseudo-enstrophy spectra. As 
discussed above, a fully developed turbulent spectrum is needed to observe 
the self-similar decay in the enstrophy.

As previously mentioned, the $G(t) \sim t^{-1/4}$ decay observed in 
the simulations is consistent with the upper limit of strict mathematical 
bounds derived for the decay of pseudo-enstrophy (see, e.g., 
\cite{Constantin1999,Cordoba2004}). From the physical point of view, 
it is interesting that such a decay law can be related with the decay 
followed by other systems that develop condensates in the inviscid 
truncated case, and large-scale coherent structures in the viscous case 
(see, e.g., \cite{Yakhot2004}). In 2D Navier-Stokes turbulence, the 
separation of scales between energy containing scales and enstrophy 
containing scales (as the former is concentrated at large scales while the 
latter is concentrated at small scales), together with assumptions of 
weak coupling between these scales \cite{Laval1999,Laval2000}, led to 
the idea that the decay of 2D Navier-Stokes turbulence can be understood 
as the superposition of a coherent (condensated) state, and a random 
phase, separated by a ``fuzzy scale'' following ideas developed for 
Bose-Einstein condensates \cite{Bogolubov1958}. In SQG, from the 
results in Sec.~\ref{sec:Theory}, we can consider a pseudo-energy 
containing scale associated with the condensation of pseudo-energy at 
low wavenumbers, which contains almost all of the system pseudo-energy, 
and a pseudo-enstrophy containing scale concentrating most of the 
pseudo-enstrophy. In the inviscid truncated case, we can easily identify 
the fuzzy scale delimiting these two phases with the thermalization 
length $L_{th} = 2\pi/k_{th}$. From Eq.~(\ref{eq:ec_q3}), dimensional 
analysis leads to
\begin{equation}
\frac{dG}{dt}= -\sigma \sim \frac{G^{3/2}}{L}
\label{eq:decay}
\end{equation}
where $L$ is a characteristic scale. Associating $L=L_{th}$, as most of 
the pseudo-enstrophy is contained below that scale, Eq.~(\ref{eq:decay}) 
is only consistent with the observed decay $G(t) \sim t^{-1/4}$ if 
$L_{th}(t) \sim t^{7/8}$. The inset in Fig.~\ref{fig:decay_todos} shows 
the time evolution of $L_{th}$ in the truncated inviscid runs with $N=1024$ 
and $N=512$ with random initial conditions, $t^{7/8}$ appears a reasonable 
fit; the other simulations are also consistent with this power law.

\section{\label{sec:Conclusions}Conclusions}

We derived statistical equilibrium solutions of the truncated inviscid SQG 
equations, and verified their validity at late times in numerical 
simulations of the truncated SQG equations. Numerical spectra are in 
agreement with the theory at all wavenumbers, although as resolution is 
increased it takes longer times for the system to reach equilibrium.

We also studied the evolution towards equilibrium of the ideal runs, 
finding that both pseudo-energy and pseudo-enstrophy spectra develop a 
viscous-like inertial range with slopes $\sim k^{-5/3}$ and $\sim k^{-8/3}$ 
respectively, and compatible with Kolmogorov-Batchelor-Kraichnan 
phenomenology. Note that in previous studies of the inviscid 
SQG equations, a steeper power law was observed \cite{Ohkitani1997}, 
but the initial conditions ${\bf I}$ that were used result in steeper 
spectra for the pseudo-energy and pseudo-enstrophy. As time evolves, the 
pseudo-enstrophy in our runs shows a gradual thermalization of higher 
wavenumber modes with $G(k) \sim k$, narrowing the viscous-like inertial 
range. The spectrum for the pseudo-energy presents a flat scaling for 
high wavenumbers and a peak at $k=1$, showing condensation of 
pseudo-energy at small wavenumbers instead of thermalization.

Defining $G_{th}$ and $E_{th}$ respectively as the pseudo-enstrophy and 
pseudo-energy contained in the thermalized modes, we studied the period 
of time during which these quantities are time-dependent. Through this 
period, the transfer of pseudo-enstrophy towards thermalized modes results 
in a viscous-like cascade with effective viscosity associated with the 
thermalized modes. This allowed us to compare spectra and decay in ideal 
runs with runs subjected to identical initial conditions but with the 
addition of dissipation. The large scales of both systems are similar 
and both systems develop a $\sim k^{-8/3}$ power law for the pseudo-energy
in accordance with the theoretical results. At later times, the power law 
is lost due to thermalization in the inviscid runs and dissipation in the 
viscous runs.

We compared the free decay of $G(t)$ with the time evolution of the 
non-thermalized $G_{nth}(t)$ in inviscid truncated runs for different 
grid sizes. Following an initial period of time in which the 
pseudo-enstrophy is quasi-conserved, viscous cases develop a 
power-law decay compatible with $G(t) \sim t^{-1/4}$ for the largest 
Reynolds numbers and spatial resolutions considered. Later in time, the 
decay becomes shallower and an exponential viscous decay of the 
remaining pseudo-enstrophy follows. The power-law decay is in 
agreement with strict mathematical bounds for the decay of the 
pseudo-enstropy \cite{Constantin1999,Cordoba2004}, and seems to be 
robust for initial conditions in which most of the pseudo-energy is 
held at the lowest modes. From the point of view of the statistical 
equilibrium solutions, the decay law can be understood as the result of 
two decaying phases weakly interacting and separated by a fuzzy scale: 
a coherent (condensated) state that contains most of the pseudo-energy, 
and a random phase that contains most of the pseudo-enstrophy.

With the exception of a small delay in the time when the quasi-conservation 
is broken, all the inviscid runs show the same evolution for the 
non-thermalized $G_{nth}(t)$ independently of the spatial resolution 
considered. The evolution of $G(t)$ approaches that of $G_{nth}$ as 
the Reynolds number is increased. Remarkably, inviscid systems as 
small as $128^2$ already give information about the free decay law 
expected for high Reynolds viscous systems.

This good agreement between the evolution of the non-thermalized components 
of the pseudo-enstrophy and the pseudo-energy in truncated inviscid runs 
at low resolution, and of the decay of pseudo-energy and pseudo-enstrophy 
in viscous runs at large resolution, further indicate a new application 
of inviscid simulations: that of estimating the decay law followed by the 
viscous system at very large Reynolds number. In this light, the decay of 
pseudo-enstrophy in SGQ turbulence in the limit of very large Reynolds 
number would be compatible with $G(t) \sim t^{-1/4}$ when all initial 
perturbations are at the gravest modes in the system. It is unclear for 
the moment whether the evolution of non-thermalized quantities in other 
inviscid truncated system can be also used as proxies of decay. This 
problem is left for future work.

\begin{acknowledgments}
The authors acknowledge support from PICT grant No. 2007-02211, UBACyT 
grant No. 20020090200692, and PIP grant No. 11220090100825.
\end{acknowledgments}

\end{document}